\documentclass[cameraready]{Interspeech}

\title{Spatial-Magnifier: Spatial upsampling for multichannel speech enhancement}

\author[affiliation={1^{\star}, 2}]{Dongheon}{Lee}
\author[affiliation={1}]{Ashutosh}{Pandey}
\author[affiliation={1}]{Sanjeel}{Parekh}
\author[affiliation={1}]{\\Daniel}{Wong}
\author[affiliation={1}]{Jacob}{Donley}
\author[affiliation={1}]{Buye}{Xu}
\author[affiliation={1^{\dagger}}]{Juan}{Azcarreta}

\address{
    $^1$ Meta Reality Labs Research \\
    $^2$ Korea Advanced Institute of Science and Technology (KAIST)
}

\email{donghen0115@gmail.com, jazcarretao@meta.com}


\usepackage{amsmath,graphicx,hyperref,amssymb, multirow, multicol, comment}
\usepackage[table,xcdraw]{xcolor}

\begin{document}

\maketitle
\begingroup
\renewcommand{\thefootnote}{}
\footnotetext{$^{\star}$ Work done during internship at Meta Reality Labs Research.}
\footnotetext{$^{\dagger}$ Corresponding author.}
\endgroup
\begin{abstract}
 While the spatial directivity of multichannel speech enhancement algorithms improves with the number of microphones, fitting large capture arrays into real-world edge devices is typically limited by physical constraints. To overcome this limitation, we propose Spatial-Magnifier, a neural network designed to generate virtual microphone (VM) signals from a limited set of real microphone (RM) measurements. Moreover, we introduce the Spatial Audio Representation Learning (SARL) framework, which leverages estimated VM signals and features to condition a downstream speech enhancement system. Experimental results demonstrate that the proposed framework outperforms existing spatial upsampling baselines across various speech extraction systems, including end-to-end multichannel speech enhancement and neural beamforming. The proposed method nearly recovers the oracle performance achieved when all microphones are available.
\end{abstract}

\section{Introduction}
\label{sec:intro}
Increasing the spatial diversity of microphone arrays by expanding the physical distance between sensors or adding more capture points can significantly boost the performance of multichannel speech enhancement (MC-SE) algorithms \cite{benesty2008microphone, VanVeen:1988, wang2020complex}. However, the spatial capture capabilities of consumer devices such as augmented reality (AR) glasses, earbuds, and hearing aids are strictly limited by physical constraints, preventing the integration of large-scale arrays.

To overcome these physical limitations, recent work has proposed neural network-based virtual microphone estimation (Neural-VME) \cite{ochiai2021neural, segawa2022neural, segawa2024neural}. In this context, a Virtual Microphone (VM) is defined as a captured signal that is available during the training phase but is absent during inference. By training a model to estimate these missing signals from a sparse set of Real Microphone (RM) measurements, Neural-VME can effectively increase the array's spatial diversity without requiring additional hardware. Previous studies have successfully applied Neural-VME to source separation by combining the estimated VM signals with RM recordings to control a mask-based beamformer \cite{ochiai2021neural, segawa2022neural, segawa2024neural}. Similarly, spatial upsampling has been utilized for Universal Acoustic Vision \cite{roman2024robust} to increase the ambisonic order by leveraging super-resolution architectures originally designed for image upscaling \cite{haris2018deep}.

Despite these advances, there has been no comprehensive study on how to condition downstream speech tasks optimally on interpolated VM signals. We argue that the primary advantage of Neural-VME lies in its ability to decouple spatial representation learning from spectral enhancement. While balancing spectral and spatial information is an intrinsic challenge in MC-SE, estimating VM signals can force the system to learn robust spatial representations that directly benefit downstream MC-SE. To maximize this potential, a generic framework applicable to the varied array geometries in consumer electronics is essential.

Notably, previous speech processing Neural-VME works have repurposed architectures originally designed for standard speech enhancement \cite{segawa2024neural, wang2024unsupervised, qiu2024transformer}. However, the distinct nature of spatial upsampling creates a pressing need for specialized models tailored to upsample microphone recordings. To this end, we introduce \textbf{Spatial-Magnifier}, a GAN-based generative network incorporating two efficient modules: a Selection Module capable of isolating the most relevant spatial features, and a Dynamic Channel Allocation (DCA) module that adaptively determines the spatial filters' importance to facilitate efficient information compression.

Furthermore, we propose \textbf{Spatial Audio Representation Learning (SARL)}, a framework that integrates Neural-VME to improve both neural beamforming and end-to-end speech enhancement. Unlike traditional virtual microphone-based Beamforming (VM-BF) \cite{segawa2024neural}, SARL can condition a downstream MC-SE model on both estimated VM signals and learned VM features. This approach enables a task we term virtual microphone-based speech enhancement (VM-SE), which improves end-to-end models directly without requiring a beamforming backend.

Extensive experiments demonstrate that the proposed approach robustly estimates VM signals across various array geometries. This expands upon existing methods that have primarily focused on linear arrays. The Spatial-Magnifier model and SARL framework achieve superior beamforming and speech extraction performance compared to conventional Neural-VME baselines. Furthermore, these performance gains are achieved at lower computational cost than existing Neural-VME baselines.

\begin{figure*}[!t]
\centering
\vspace{-2em}
\includegraphics[width=6.5in]{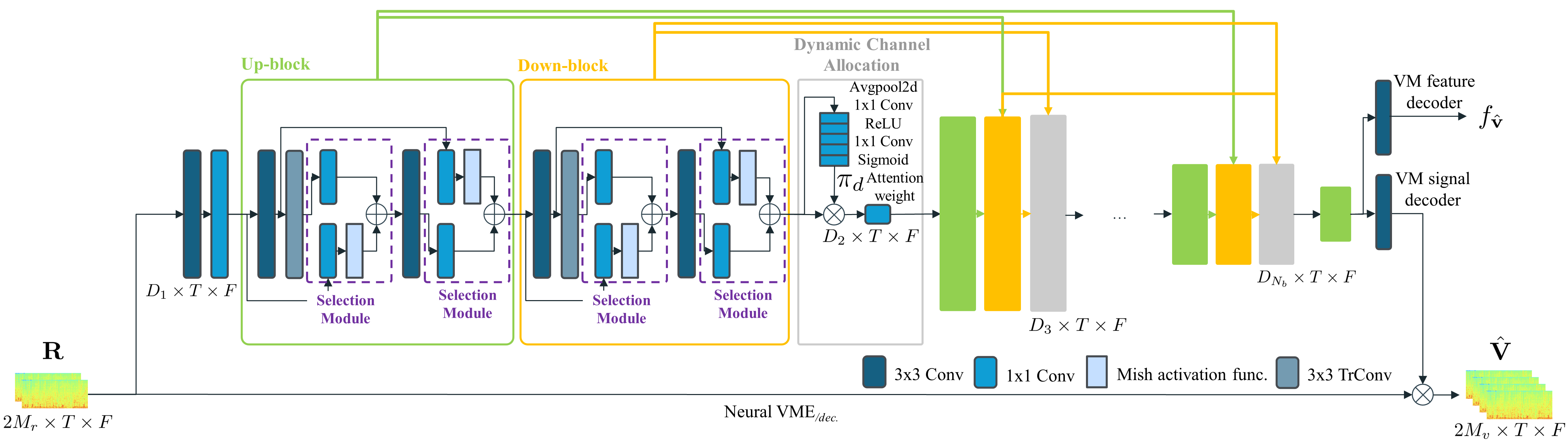}
\vspace{-1em}
\caption{Architecture of the Spatial-Magnifier generator. The network jointly generates VM signals and VM features.}
\label{fig:model}
\vspace{-2em}
\end{figure*}

\section{Proposed method}
\label{sec:method}
\subsection{Mathematical modeling of neural beamforming}
MC-SE is the task of estimating a direct-path speech signal $\mathbf{x}_{ref} \in \mathbb{R}^{1 \times N}$ given multichannel noisy speech $\mathbf{y} \in \mathbb{R}^{M \times N}$ consisting of $M$ channels and $N$ samples, which can be expressed as
\begin{align}
\begin{aligned}
\mathbf{y} &= \mathbf{x} + \mathbf{x}_{rev} + \mathbf{n},
\end{aligned}
\end{align}
where $\mathbf{x} \in \mathbb{R}^{M \times N}$, $\mathbf{x}_{rev} \in \mathbb{R}^{M \times N}$, and $\mathbf{n} \in \mathbb{R}^{M \times N}$ denote the multichannel waveforms of the direct-path speech, its reverberation, and additive noise, respectively. The target signal $\mathbf{x}_{ref}$ corresponds to a selected reference channel from $\mathbf{x}$.

We utilize a discriminative multichannel neural network to estimate the target signal at the reference microphone:
\begin{equation}
\begin{aligned}
\label{eq:mc_se}
\hat{\mathbf{x}}^{se}_{ref}=\text{MC-SE}(\mathbf{y}).
\end{aligned}
\end{equation}
In our framework, we leverage the estimated $\hat{\mathbf{x}}^{se}_{ref}$ as an approximation of the target signal to derive adaptive coefficients $\boldsymbol{W} \in \mathbb{C}^{M \times T \times F}$ for frequency-domain beamforming:
\begin{equation}
\label{eq:vm_bf}
\begin{aligned}
\hat{\boldsymbol{X}}_{ref}^{bf} &= \boldsymbol{W}^{\mathsf{H}} \boldsymbol{Y},
\end{aligned}
\end{equation}
where $\boldsymbol{Y} \in \mathbb{C}^{M \times T \times F}$ denotes the STFT of the input signal across $M$ microphones, $T$ frames, and $F$ frequency bins. $(\cdot)^{\mathsf{H}}$ denotes the Hermitian operator. The output $\hat{\boldsymbol{X}}_{ref}^{bf} \in \mathbb{C}^{T \times F}$ is the frequency-domain estimated target signal. The weights $\boldsymbol{W}$ can be calculated in closed-form using classical beamformers such as the multichannel Wiener filter (MCWF) \cite{benesty2008microphone} or the minimum variance distortionless response (MVDR) \cite{souden2009optimal}. We utilize a time-varying filter with windowed mean processing to ensure stability. Linear filtering effectively mitigates non-linear distortions introduced by neural networks, which can be beneficial for multi-stage processing \cite{pandey2025ultra, wang2022stft, wang2020derev} and to enhance the performance of ASR applications \cite{jahn2016}.

\subsection{Neural-VME for speech enhancement}
Neural-VME leverages a neural network to estimate missing Virtual Microphone (VM) signals from a sparse set of Real Microphone (RM) measurements. During training, all microphone signals $\mathbf{y}=[\mathbf{r}, \mathbf{v}]$ are available, where $\mathbf{r} \in \mathbb{R}^{M_r \times N}$ and $\mathbf{v} \in \mathbb{R}^{M_v \times N}$ denote the $M_r$ real and $M_v$ virtual channels, respectively. Hence, the task of Neural-VME is to produce an estimate $\hat{\mathbf{v}} \in \mathbb{R}^{M_v \times N}$ of the VM signals  given $\mathbf{r}$:
\begin{equation}
\label{eq:neural-vme}
\hat{\mathbf{v}} = \text{Neural-VME}(\mathbf{r}).
\end{equation}
We leverage virtual microphone-based beamforming (VM-BF) \cite{segawa2024neural}, where the augmented signal $\bar{\mathbf{y}}=[\mathbf{r}, \hat{\mathbf{v}}] \in \mathbb{R}^{M \times N}$ (with $M = M_r + M_v$) is used to calculate the Spatial Covariance Matrices (SCM)  for a beamforming back-end \cite{benesty2008microphone} to derive the filters $\boldsymbol{W}$, which are then applied to $\bar{\boldsymbol{Y}}$ following Eq. \ref{eq:vm_bf}. VM-BF jointly optimizes a multi-task objective for Neural-VME and beamforming, learning how to generate VM signals that aim to increase the numerical rank of the SCM from $M_r$ to $M$ for SE. 

\subsection{Spatial-Magnifier model}
The proposed Spatial-Magnifier model is a generative adversarial network (GAN) \cite{goodfellow2014generative} that consists of convolutions designed to exploit inter-channel relationships. Figure \ref{fig:model} depicts the architecture of the Spatial-Magnifier generator, which is inspired by the deep back-projection network (DBPN) for image super-resolution \cite{haris2018deep}. Spatial-Magnifier processes the RM signals $\boldsymbol{R} \in \mathbb{C}^{M_r \times T \times F}$ in the frequency-domain by treating the microphone indices as the channel dimension and concatenating real and imaginary components. An initial 2D convolution expands the input $2 \times M_r$ to $D_1$ channels. Features then undergo $N_b$ stages of alternating up-blocks, down-blocks, and our proposed DCA modules. The DCA module utilizes dynamic convolutions \cite{chen2020dynamic} to compute channel-wise attention scores, weighting a pointwise convolution that adaptively reduces dimensionality from $D_1$ to $D_2$ for efficient compression.

Previously, up-blocks and down-blocks utilized simple addition and subtraction, applying identical operations across all channels, which limited their flexibility. To address this, we introduce a selection module (SM) that incorporates pointwise convolution followed by Mish activation \cite{misra2019mish} to form a gating mechanism \cite{lee2025deft} before the addition operation. This approach extracts features channel-wise adaptively, enhancing performance with minimal computational overhead. Since DBPN architectures resemble dense blocks, they often incur high computational costs. Given that Neural-VME targets real-world devices, maximizing performance gains while minimizing computational load is essential. For additional efficiency, group convolution is employed in the down-blocks. We adopt the discriminator from the conformer-based MetricGAN (CMGAN) \cite{abdulatif2024cmgan}.

\begin{figure}[!b]
\centering
\vspace{-1.5em}
\includegraphics[width=3.15in]{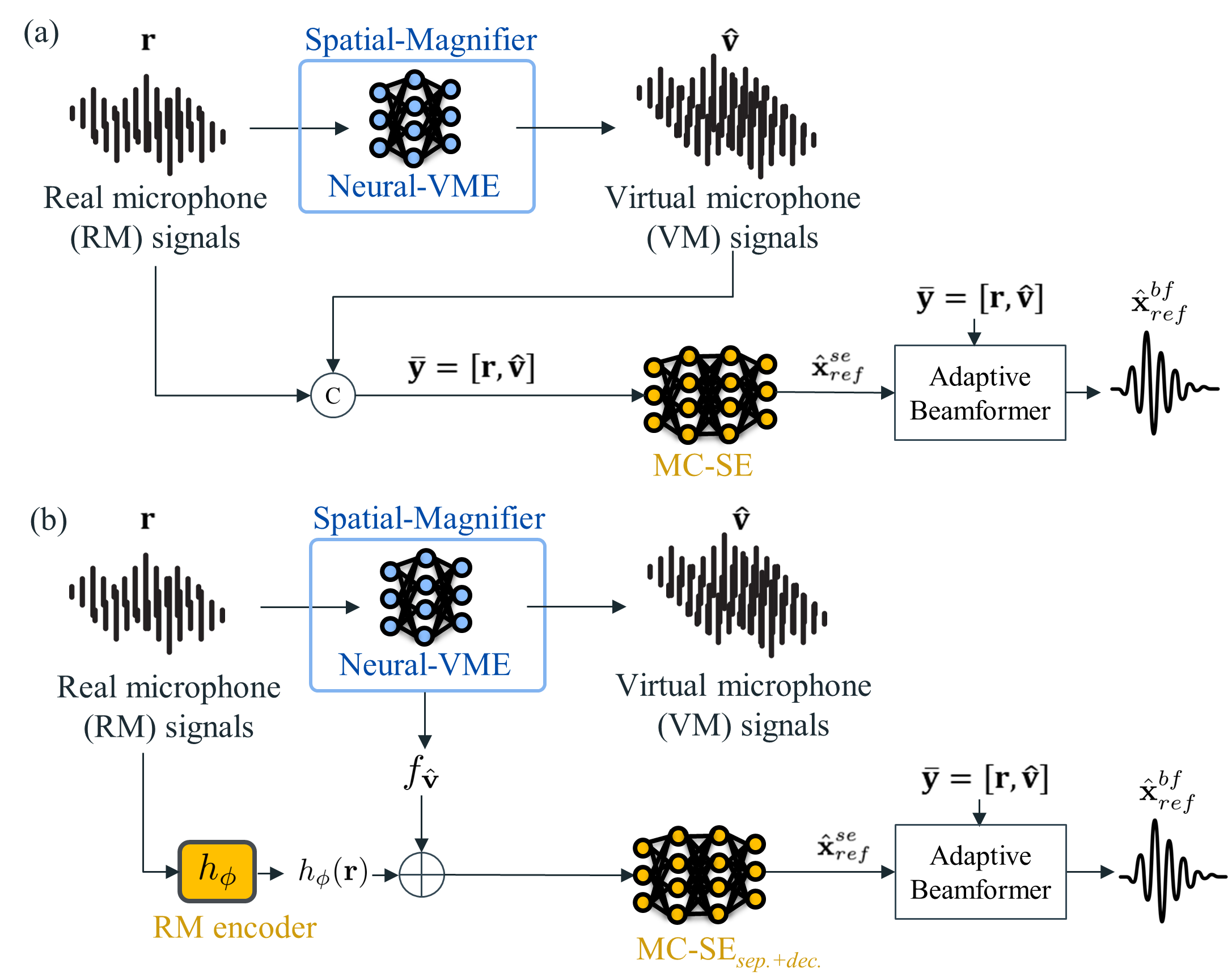}
\vspace{-2em}
\caption{Overall framework of Spatial Audio Representation Learning (SARL): (a) SARL-Signal and (b) SARL-Feature frameworks. Spatial-Magnifier serves as the Neural-VME model, while SARL represents the conditioning method for the MC-SE model.}
\label{fig:training_method}
\vspace{-2em}
\end{figure}

\subsection{Spatial Audio Representation Learning}
We propose Spatial Audio Representation Learning (SARL) to condition the MC-SE model on the estimated VM signals. As illustrated in Figure \ref{fig:training_method}, SARL encompasses two paradigms: SARL-signal (SARL-S) and SARL-feature (SARL-F). Both strategies aim to optimize the enhanced signal $\hat{\mathbf{x}}^{se}_{ref}$ by augmenting the RM observations with virtual spatial information, a task that we call virtual microphone-based speech enhancement (VM-SE). Within the SARL framework, we utilize a pre-trained MC-SE model originally trained with the overall signals $\mathbf{y}$. We then fine-tune this model while training the Neural-VME model from scratch, maintaining the same computational cost during inference. VM-SE improves end-to-end MC-SE performance without relying on a beamformer back-end. 

\subsubsection{SARL-S: Signal-Level Augmentation}
SARL-S is a direct spatial upsampling approach where the Spatial-Magnifier estimates explicit VM signals that are concatenated with the RM signals to form the augmented signal $\bar{\mathbf{y}} = [\mathbf{r}, \hat{\mathbf{v}}]$. This augmented signal is then directly processed by an MC-SE model as in Equation \ref{eq:mc_se}. By providing raw waveforms, SARL-S allows the downstream model to utilize improved spatial information across the expanded array geometry.

\subsubsection{SARL-F: Feature-Level Augmentation}
In contrast, SARL-F operates in a latent space to provide robust conditioning. Since common MC-SE models can be decomposed into
an encoder-separator-decoder topology \cite{quan2024spatialnet, luo2019conv, lee2024deftan}, by defining the encoder as $h_{\phi}(\cdot)$, and the separator+decoder modules as $\text{MC-SE}_{\textit{sep.}+\textit{dec.}}(\cdot)$, the enhanced signal is given by:
\begin{align}
\hat{\mathbf{x}}^{se_{\bar{\mathbf{y}}}} &= \text{MC-SE}_{\textit{sep.}+\textit{dec.}}(h_{\phi}(\mathbf{r}) + f_{\hat{\mathbf{v}}}),
\end{align}
where $f_{\hat{\mathbf{v}}} \in \mathbb{R}^{H \times T \times F}$ denotes the estimated VM features by Spatial-Magnifier, where $H$ is the embeddings size. In SARL-F, Spatial-Magnifier estimates representations equivalent to an encoded spatial embedding, which are fused with the encoded RM signals $h_{\phi}(\mathbf{r}) \in \mathbb{R}^{H \times T \times F}$ via element-wise addition \cite{yang2024self, hong2025efficient}. This latent fusion allows the separator to exploit spatial diversity even when the raw VM waveform reconstruction is challenging, acting as a high-level spatial regularizer. 


\section{Experiments}
\label{sec:exp}

\subsection{Datasets}
We used the Interspeech 2020 DNS challenge speech and noise corpora \cite{reddy2020interspeech} to simulate 50,000, 2,000, and 3,000 clips of 10 s duration for training, validation, and testing, respectively. Spatial data were simulated via \textit{Pyroomacoustics} \cite{scheibler2018pyroomacoustics} using the image source method with an order of six. The six-channel array consisted of a four-channel circular array with a radius of 10 cm and two vertical microphones placed 10 cm above and below the center. The length, width, and height of the room were uniformly distributed within [3, 10], [3, 10], and [2, 5] m, respectively, with an absorption coefficient sampled from the range [0.1, 0.5], resulting in reverberation time (RT60) in the range [0.15, 1.75] s. The signal-to-noise ratio (SNR) and signal-to-interference ratio (SIR) were sampled within [$-$10, 5] dB, with sources placed [0.5, 2.5] m from the array center. The experiments covered both conventional omnidirectional SE (omni-SE) and Field-of-View SE (FoV-SE) tasks \cite{xufovnet}. For FoV-SE, the target was within $\pm 20^{\circ}$ (azimuth and elevation) relative to the front axis. Up to four interfering talkers were placed outside this FoV area. The number of babble talkers and noise sources ranged from 0 to 10 for omni-SE and from 0 to 5 for FoV-SE.

\subsection{Experimental setup}
We computed the short-time Fourier transform (STFT) with a 16 ms square-root Hanning window, an 8 ms hop size, and a 16 kHz sampling rate. Time-varying beamformer weights were computed block-wise using a 25-frame window \cite{wang2022mf}. For the Spatial-Magnifier, we set $N_b=5$ and channel dimensions $[D_1, \dots, D_5] = [128, 96, 64, 48, 32]$. The loss function combined time-domain SNR losses for Neural-VME and VM-BF, along with adversarial losses \cite{kong2020hifi} for the generator and discriminator with weights of 0.3:0.7:0.01:0.01, respectively. The first RM channel is utilized as the target reference signal. The model was trained using the Adam optimizer with a learning rate of 0.001 for 100 epochs, with a batch size of 64 across 32 H100 GPUs. Performance was evaluated using SI-SDR \cite{le2019sdr}, SNR, narrowband PESQ \cite{rix2001perceptual}, and STOI \cite{taal2010short}.

\begin{table}[!b]
\vspace{-1.5em}
\caption{Ablation study on training methods, RM: 2ch, VM: 4ch\label{tab:ablation}}
\vspace{-1em}
\centering
\setlength{\tabcolsep}{1.5pt}
\scalebox{0.78}{
\begin{tabular}{c|l|cc|cccc}
\hline
\multirow{2}{*}{Model type} & \multirow{2}{3.15cm}{Training method} & \multicolumn{2}{c|}{Neural-VME} & \multicolumn{4}{c}{VM-BF} \\ \cline{3-8} 
 &  & SI-SDR & SNR & SI-SDR & SNR & PESQ & STOI \\ \hline
\multicolumn{2}{c|}{unprocessed} & - & - & -11.0 & -9.97 & 1.29 & 50.1 \\ \hline
\multicolumn{2}{c|}{SpatialNet + MCWF 2ch} & - & - & 2.19 & 4.57 & 1.97 & 70.4 \\ \hline
\multirow{8}{*}{\begin{tabular}[c]{@{}c@{}}Spatial-\\ Magnifier\end{tabular}} 
 & Neural-VME (freeze) & 3.55 & 5.27 & 4.01 & 5.71 & 2.08 & 75.1 \\
 & Neural-VME (unfreeze) & 3.45 & 5.20 & 5.30 & 6.71 & 2.14 & 76.9 \\
 & \cellcolor{gray!30}\textbf{SARL-F} & \cellcolor{gray!30}3.45 & \cellcolor{gray!30}5.20 & \cellcolor{gray!30}6.10 & \cellcolor{gray!30}7.27 & \cellcolor{gray!30}2.33 & \cellcolor{gray!30}80.4 \\
 & \quad- w/o VM loss & - & - & 5.29 & 6.68 & 2.21 & 77.9 \\
 & \quad- w/o VM signals & \textbf{3.54} & \textbf{5.27} & 2.74 & 4.87 & 2.02 & 72.1 \\
 & \cellcolor{gray!30}\textbf{SARL-S} & \cellcolor{gray!30}3.44 & \cellcolor{gray!30}5.20 & \cellcolor{gray!30}\textbf{7.10} & \cellcolor{gray!30}\textbf{8.09} & \cellcolor{gray!30}\textbf{2.40} & \cellcolor{gray!30}\textbf{82.1} \\
 & \quad- w/o VM loss & - & - & 6.89 & 7.91 & 2.39 & 81.9 \\
 & \quad- w/o VM signals & 3.65 & 5.34 & 3.12 & 5.12 & 2.04 & 73.3 \\ \hline
\multicolumn{2}{c|}{SpatialNet + MCWF 6ch} & - & - & 8.35 & 9.06 & 2.41 & 84.6 \\ \hline
\end{tabular}}
\end{table}

\begin{table}[!b]
\vspace{-0.5em}
\caption{Ablation on Spatial-Magnifier, RM: 2ch, VM: 4ch\label{tab:ablation_model}}
\vspace{-1em}
\centering
\setlength{\tabcolsep}{1.5pt}
\scalebox{0.78}{
\begin{tabular}{c|l|cc|cccc}
\hline
\multirow{2}{*}{\begin{tabular}[c]{@{}c@{}}Training\\ method\end{tabular}} & \multirow{2}{*}{Model type} & \multicolumn{2}{c|}{Neural-VME} & \multicolumn{4}{c}{VM-BF} \\ \cline{3-8}
 &  & SI-SDR & SNR & SI-SDR & SNR & PESQ & STOI \\ \hline
\multicolumn{2}{c|}{SpatialNet + MCWF 2ch} & - & - & 2.19 & 4.57 & 1.97 & 70.4 \\ \hline
\multirow{4}{1.4cm}{\centering SARL-F}
 & \cellcolor{gray!30}\textbf{Spatial-Magnifier} & \cellcolor{gray!30}3.45 & \cellcolor{gray!30}5.20 & \cellcolor{gray!30}6.10 & \cellcolor{gray!30}7.27 & \cellcolor{gray!30}\textbf{2.33} & \cellcolor{gray!30}80.4 \\ 
 & \quad- w/o GAN & 3.47 & 5.21 & 6.27 & 7.40 & 2.33 & 80.6 \\
 & \quad- w/o selection module & 3.39 & 5.16 & 5.98 & 7.18 & 2.30 & 79.7 \\
 & \quad- w/o DCA & 3.40 & 5.17 & 5.54 & 6.87 & 2.16 & 76.9 \\ \hline
\multirow{4}{1.4cm}{\centering SARL-S}
 & \cellcolor{gray!30}\textbf{Spatial-Magnifier} & \cellcolor{gray!30}3.44 & \cellcolor{gray!30}5.20 & \cellcolor{gray!30}\textbf{7.10} & \cellcolor{gray!30}\textbf{8.09} & \cellcolor{gray!30}\textbf{2.40} & \cellcolor{gray!30}\textbf{82.1} \\ 
 & \quad- w/o GAN & \textbf{3.49} & \textbf{5.23} & 7.06 & 8.06 & 2.39 & 81.8 \\
 & \quad- w/o selection module & 3.39 & 5.16 & 6.82 & 7.85 & 2.35 & 81.5 \\
 & \quad- w/o DCA & 3.41 & 5.16 & 7.01 & 8.00 & 2.38 & 81.9 \\ \hline
\multicolumn{2}{c|}{SpatialNet + MCWF 6ch} & - & - & 8.35 & 9.06 & 2.41 & 84.6 \\ \hline
\end{tabular}}
\vspace{-2em}
\end{table}

\begin{table*}[!t]
\vspace{-1.0em}
\caption{VM-BF comparison against baseline models \label{tab:comp}}
\vspace{-1.0em}
\centering
\scalebox{0.78}{
\setlength{\tabcolsep}{5pt}
\begin{tabular}{l|cc|cccc|cc|cccc|cc}
\hline  & 
  \multicolumn{6}{c|}{RM: 2ch, VM: 1ch} & \multicolumn{6}{c|}{RM: 2ch, VM: 4ch} & 
  \multirow{3}{*}{Param.} &
  \multirow{3}{*}{MAC/s} \\ \cline{2-13}
 &
  \multicolumn{2}{c|}{Neural-VME} &
  \multicolumn{4}{c|}{VM-BF} & 
  \multicolumn{2}{c|}{Neural-VME} &
  \multicolumn{4}{c|}{VM-BF} & & \\ \cline{2-7} \cline{8-13}
                          & SI-SDR & SNR  & SI-SDR & SNR  & PESQ & STOI & SI-SDR & SNR  & SI-SDR & SNR  & PESQ & STOI &        &         \\ \hline
SpatialNet + MCWF 2ch &  - & - & 3.14 & 4.96 & 2.13 & 75.5 & - & - & 3.14 & 4.96 & 2.13 & 75.5 & 1.2 M & 19.8 G \\ \hline
\quad + MC Conv-TasNet (STL) \cite{ochiai2021neural} & 2.85 & 4.81 & 3.37 & 5.10 & 2.14 & 76.1 & 2.84 & 4.80 & 3.69 & 5.31 & 2.16 & 76.8 & +13.0 M & +20.5 G \\
\quad + MC Conv-TasNet (MTL) \cite{segawa2024neural} &  2.83 & 4.79 & 3.78 & 5.37 & 2.17 & 76.9 & 2.76 & 4.75 & 4.89 & 6.16 & 2.24 & 79.3 & +13.0 M & +20.5 G \\
\quad + SpatialNet-VME &  \textbf{2.90} & \textbf{4.84} & 4.80 & 5.39 & 2.17 & 76.9 & 2.40 & 4.50 & 4.87 & 6.15 & 2.23 & 79.2 & +1.2 M & +19.8 G \\
\quad + Spatial-Magnifier (VME) &  2.77 & 4.76 & 5.58 & 6.69 & 2.31 & 80.6 & \textbf{2.89} & \textbf{4.83} & 5.84 & 6.88 & 2.36 & 81.6 & \textbf{+1.2 M} & \textbf{+19.2 G} \\
\quad \cellcolor{gray!30}\textbf{+ Spatial-Magnifier (SARL-F)} & \cellcolor{gray!30}2.61 & \cellcolor{gray!30}4.66 & \cellcolor{gray!30}6.32 & \cellcolor{gray!30}7.27 & \cellcolor{gray!30}2.36 & \cellcolor{gray!30}82.4 & \cellcolor{gray!30}2.78 & \cellcolor{gray!30}4.76 & \cellcolor{gray!30}7.72 & \cellcolor{gray!30}8.37 & \cellcolor{gray!30}2.51 & \cellcolor{gray!30}85.1 & \cellcolor{gray!30}+1.5 M & \cellcolor{gray!30}+24.4 G \\ 
\quad \cellcolor{gray!30}\textbf{+ Spatial-Magnifier (SARL-S)} &  \cellcolor{gray!30}2.69 & \cellcolor{gray!30}4.70 & \cellcolor{gray!30}\textbf{6.87} & \cellcolor{gray!30}\textbf{7.70} & \cellcolor{gray!30}\textbf{2.40} & \cellcolor{gray!30}\textbf{83.1} & \cellcolor{gray!30}2.78 & \cellcolor{gray!30}4.76 & \cellcolor{gray!30}\textbf{8.37} & \cellcolor{gray!30}\textbf{8.98} & \cellcolor{gray!30}\textbf{2.57} & \cellcolor{gray!30}\textbf{86.5} & \cellcolor{gray!30}\textbf{+1.2 M} & \cellcolor{gray!30}\textbf{+19.2 G} \\ \hline
SpatialNet + MCWF 3/6 ch &  - & - & 5.41 & 6.57 & 2.25 & 80.6 & - & - & 9.49 & 9.91 & 2.57 & 88.9 & 1.2 M & 19.8 G  \\ \hline
Oracle MCWF 3/6 ch & - & - & 6.65 & 7.55 & 2.41 & 84.6 & - & - & 11.78 & 12.06 & 2.70 & 92.4 & - & - \\ \hline
\end{tabular}}
\vspace{-1.5em}
\end{table*}

\section{Results}
\label{sec:results}
For the ablation study and baseline comparison, we employed SpatialNet-small \cite{quan2024spatialnet} as the MC-SE model combined with an MCWF \cite{benesty2008microphone, wang2022stft} beamformer. Neural network computation load is reported as Multiply Accumulates per second (MAC/s).

\subsection{Ablation study}
This analysis focuses on the FoV-SE task suitable for MC-SE. In Table \ref{tab:ablation}, we first show that while joint VM-BF and Neural-VME fine-tuning of a pre-trained MC-SE model improved VM-BF performance, it remained inferior to the proposed SARL methods. This demonstrates that conditioning the MC-SE model directly on VM features outperforms standard fine-tuning. Next, removing the VM loss drops the performance of both SARL methods. The performance drop without VM loss confirms the necessity of virtual spatial information, suggesting that leveraging generated spatial information for VM-BF is crucial. Notably, even when excluding the VM signals from the adaptive beamforming, the VM-BF improves with respect to the Spatialnet+MCWF system that utilizes only 2ch-RM, which suggests the effectiveness of SARL conditioning.

We report the Spatial-Magnifier architecture ablation study in Table \ref{tab:ablation_model}. While GAN provides the highest Neural-VME performance, its effectiveness for VM-BF seems modest. In contrast, VM-BF performance degrades significantly without the selection or DCA modules. Both modules are highly efficient, each adding only 0.1M parameters and 0.1 GMAC/s.

The results for the selection module suggest that weighted sums per convolutional channel enhance the flexibility of spatial information utilization. Similarly, the performance gain from DCA reveals that attention scores play a crucial role in effectively compressing spatial information. 


\subsection{Comparison with existing Neural-VME models}
For comparisons with previous work \cite{segawa2024neural} in the omni-SE task, Table \ref{tab:comp} shows that simply employing a high-performance MC-SE model is not the optimal approach for VM-BF. We confirm this finding by also utilizing SpatialNet \cite{quan2024spatialnet} as an architecture for the Neural-VME task. Overall, the proposed Spatial-Magnifier achieves superior VM-BF results with lower computational cost. When estimating multiple VM signals Spatial-Magnifier outperforms other baselines also in the Neural-VME task, highlighting that a specialized network design exploiting spatial information across the channel dimension is critical for spatial upsampling. Also, the SARL training framework enables joint optimization of Neural-VME accuracy while concurrently achieving the highest VM-BF performance through learned spatial audio representations. 


Interestingly, the 2ch-RM/1ch-VM SpatialNet+MCWF with SARL outperforms the 3ch-RM SpatialNet+MCWF, proving the joint multi-task loss creates effective spatial representations for downstream enhancement. Furthermore, the 2ch-RM/1ch-VM SARL-S configuration synthesizes virtual channels with non-linear spatial priors, acting as an optimized spatial regularizer that achieves better noise suppression than the 3ch-RM oracle MCWF. However, all the 2ch-RM/4ch-VM setup still trail the 6ch-RM oracle MCWF, indicating room for further improvement in complex spatial upsampling scenarios.

\subsection{Versatility across various processing strategies}
The performance across variants involving different processing strategies is depicted in Table \ref{tab:general} for the FoV-SE task. To evaluate the versatility of our approach, we expanded the experiments to include a challenging 2ch-RM/8ch-VM scenario. We also assessed the reliability on core processing components by adopting a mask-based Souden MVDR \cite{souden2009optimal} as the adaptive beamformer and switching the MC-SE model to a multichannel recurrent neural network (MC-RNN) \cite{pandey2023simple}. Furthermore, we validated the method on a 7-channel array simulated with measured Array Transfer Functions (ATFs) from an array comprising 5 microphones mounted in smart glasses (RM) and 2 channels representing HRTF responses (VM).

In the challenging 2ch-RM/8ch-VM configuration, the model achieved performance near that of a physical 10-channel system, indicating it generates substantial spatial information for VM-BF even from limited data. The framework's robustness across different back-ends was demonstrated by switching from MCWF to MVDR while maintaining competitive results. Similarly, replacing the backbone MC-SE model with MC-RNN preserved performance gains, confirming the architecture-agnostic nature of the approach. Finally, the method achieved results comparable to a 7ch-RM model on the smart glasses form-factor with HRTF recordings, suggesting broad applicability to diverse real-world array geometries. However, as Neural-VME is intrinsically linked to the training array geometry, generating signals at arbitrary positions is difficult.

Finally, we verified whether Neural-VME could augment a state-of-the-art end-to-end model, such as SpatialNet \cite{quan2024spatialnet}. By performing the VM-SE task using the combination of SpatialNet-small and the proposed Spatial-Magnifier, our approach achieved a higher speech quality than SpatialNet-large 2ch-RM, despite the significantly lower computational costs of our configuration (parameter size: 2.7M vs. 6.5M, computational complexity: 44.2 GMAC/s vs. 110 GMAC/s). This suggests that when the number of microphones is constrained, leveraging virtual spatial information is a more effective strategy for enhancing performance than simply increasing the model size. Ultimately, using VM signals as training targets forces the model to learn rich spatial priors, improving resolution and leading to superior speech enhancement performance.

\begin{table}[!t]
\vspace{0.5em}
\caption{Variants involving different processing strategies \label{tab:general}}
\vspace{-1.0em}
\centering
\setlength{\tabcolsep}{1.5pt}
\scalebox{0.78}{
\begin{tabular}{c|l|cc|cccc}
\hline
 Variant &  & \multicolumn{2}{c|}{Neural-VME} & \multicolumn{4}{c}{VM-BF (or VM-SE)} \\ \cline{3-8}
 types & & SI-SDR & SNR & SI-SDR & SNR & PESQ & STOI \\ \hline
& SpatialNet + MCWF 2ch & - & - & 2.19 & 4.57 & 1.97 & 70.4 \\ \cline{2-8}
& \cellcolor{gray!30}\textbf{\quad + SARL-F} & \cellcolor{gray!30}5.51 & \cellcolor{gray!30}6.71 & \cellcolor{gray!30}6.59 & \cellcolor{gray!30}7.62 & \cellcolor{gray!30}2.37 & \cellcolor{gray!30}81.6 \\
 & \cellcolor{gray!30}\textbf{\quad + SARL-S} & \cellcolor{gray!30}\textbf{5.57} & \cellcolor{gray!30}\textbf{6.75} & \cellcolor{gray!30}\textbf{7.06} & \cellcolor{gray!30}\textbf{8.05} & \cellcolor{gray!30}\textbf{2.40} & \cellcolor{gray!30}\textbf{82.4} \\ \cline{2-8}
\multirow{-4}{*}{VM 8ch} & SpatialNet + MCWF 10ch & - & - & 9.56 & 10.10 & 2.56 & 88.3 \\ \hline
  
& SpatialNet + MVDR 2ch & - & - & 3.07 & 5.09 & 2.11 & 74.6 \\ \cline{2-8}
 & \cellcolor{gray!30}\textbf{\quad + SARL-F} & \cellcolor{gray!30}\textbf{3.45} & \cellcolor{gray!30}\textbf{5.20} & \cellcolor{gray!30}\textbf{6.72} & \cellcolor{gray!30}\textbf{7.75} & \cellcolor{gray!30}\textbf{2.39} & \cellcolor{gray!30}\textbf{81.7} \\
& \cellcolor{gray!30}\textbf{\quad + SARL-S} & \cellcolor{gray!30}3.37 & \cellcolor{gray!30}5.14 & \cellcolor{gray!30}6.32 & \cellcolor{gray!30}7.45 & \cellcolor{gray!30}2.35 & \cellcolor{gray!30}80.6 \\ \cline{2-8}
\multirow{-4}{*}{MVDR} & SpatialNet + MVDR 6ch & - & - & 8.03 & 8.78 & 2.52 & 85.2 \\ \hline

& MC-RNN + MCWF 2ch & - & - & -2.66 & 2.38 & 1.67 & 59.4 \\ \cline{2-8}
 & \cellcolor{gray!30}\textbf{\quad + SARL-F} & \cellcolor{gray!30}\textbf{3.54} & \cellcolor{gray!30}\textbf{5.26} & \cellcolor{gray!30}-1.31 & \cellcolor{gray!30}3.02 & \cellcolor{gray!30}1.80 & \cellcolor{gray!30}64.7 \\
& \cellcolor{gray!30}\textbf{\quad + SARL-S} & \cellcolor{gray!30}3.50 & \cellcolor{gray!30}5.24 & \cellcolor{gray!30}\textbf{1.15} & \cellcolor{gray!30}\textbf{4.17} & \cellcolor{gray!30}\textbf{1.99} & \cellcolor{gray!30}\textbf{70.3} \\ \cline{2-8}
\multirow{-4}{*}{\begin{tabular}[c]{@{}c@{}}MC-\\ RNN\end{tabular}} & MC-RNN + MCWF 6ch & - & - & 2.79 & 4.95 & 2.01 & 72.3 \\ \hline

& SpatialNet + MCWF 3ch & - & - & 2.48 & 4.83 & 1.92 & 72.6 \\ \cline{2-8}
& \cellcolor{gray!30}\textbf{\quad + SARL-F} & \cellcolor{gray!30}3.97 & \cellcolor{gray!30}5.56 & \cellcolor{gray!30}4.97 & \cellcolor{gray!30}6.48 & \cellcolor{gray!30}2.10 & \cellcolor{gray!30}79.1 \\
 & \cellcolor{gray!30}\textbf{\quad + SARL-S} & \cellcolor{gray!30}\textbf{4.31} & \cellcolor{gray!30}\textbf{5.80} & \cellcolor{gray!30}\textbf{5.90} & \cellcolor{gray!30}\textbf{7.22} & \cellcolor{gray!30}\textbf{2.28} & \cellcolor{gray!30}\textbf{82.1} \\ \cline{2-8}
\multirow{-4}{*}{\begin{tabular}[c]{@{}c@{}}Smart\\ glasses\end{tabular}} & SpatialNet + MCWF 7ch & - & - & 7.34 & 8.26 & 2.36 & 85.9 \\ \hline

& SpatialNet-small 2ch & - & - & 8.16 & 8.99 & 2.62 & 86.2 \\ \cline{2-8}
& \cellcolor{gray!30}\textbf{\quad + SARL-F} & \cellcolor{gray!30}3.54 & \cellcolor{gray!30}5.26 & \cellcolor{gray!30}9.04 & \cellcolor{gray!30}9.73 & \cellcolor{gray!30}\textbf{2.72} & \cellcolor{gray!30}\textbf{87.6} \\
 & \cellcolor{gray!30}\textbf{\quad + SARL-S} & \cellcolor{gray!30}\textbf{3.58} & \cellcolor{gray!30}\textbf{5.29} & \cellcolor{gray!30}8.80 & \cellcolor{gray!30}9.43 & \cellcolor{gray!30}2.62 & \cellcolor{gray!30}86.5 \\
& SpatialNet-large 2ch & - & - & \textbf{9.33} & \textbf{9.93} & 2.62 & 87.5 \\ \cline{2-8}
\multirow{-5}{*}{VM-SE} & SpatialNet-small 6ch & - & - & 12.1 & 12.4 & 2.92 & 92.3 \\ \hline
  
\end{tabular}}
\vspace{-1.5em}
\end{table}

\section{Conclusion}
\label{sec:conclusion}
This paper introduces Spatial-Magnifier, a dedicated network for audio spatial upsampling, and SARL, a novel training framework for virtual microphone-based beamforming (VM-BF) and speech enhancement (VM-SE). The proposed method achieves high VM-BF performance by effectively leveraging spatial information to estimate multiple VM representations to condition a downstream task. Furthermore, the method showcases robustness across various speech enhancement tasks, array geometries, and downstream model architectures.

\section{Generative AI Use Disclosure}
Generative AI tools (Gemini, ChatGPT) were used for editing and polishing the manuscript. All scientific content, experimental design, and results were produced by the authors.

\bibliographystyle{IEEEtran}
\bibliography{mybib}

\end{document}